\documentclass[prl,10pt,twocolumn,showpacs,superscriptaddress,balancelastpage]{revtex4}

\input{tcilatex}

\begin{document}

\title{Quantum Hall Effect in a Rotating Bose-Einstein Condensate: An Atomic
Twin of the Electronic Brother?}
\author{Zeng-Bing Chen}
\email{zbchen@ustc.edu.cn}
\author{Bo Zhao}
\affiliation{Department of Modern Physics, University of Science and Technology of China,
Hefei, Anhui 230027, China}
\author{Yong-De Zhang}
\affiliation{CCAST (World Laboratory), P.O. Box 8730, Beijing 100080, China and
Department of Modern Physics, University of Science and Technology of China,
Hefei, Anhui 230027, China}
\date{\today }

\begin{abstract}
We exploit the analogy with the quantum Hall (QH) effect for electrons to
study the possible atomic QH states of a rapidly-rotating Bose-Einstein
condensate. Actually, there is a nearly perfect map of the present problem
in the QH regime to the QH physics for electrons. The profound map enables
one to give a physically appealing definitions of the filling fraction and
the \textquotedblleft atomic Hall conductance\textquotedblright\ that is
quantized for atomic Laughlin states. This quantization might imply \textit{%
an exotic fractionalization of atomic mass}. We also briefly discuss an
effective Chern-Simons theory for describing the atomic QH liquids where a
gravitational-like field naturally emerges.
\end{abstract}

\maketitle

Bose-Einstein condensates (BECs) in dilute systems of trapped neutral atoms 
\cite{BEC,RMP} have offered a fascinating testing ground for some basic
concepts in elementary quantum mechanics and quantum many-body theory as
well as for searching new macroscopic quantum coherent phenomena. A unique
feature of quantum degenerate atomic gases is that they can be easily
controlled and manipulated by electromagnetic fields. The recent
observations \cite{vortex-exp,vortex-Science,vortex-critical,vortex-topo} of
large vortex arrays in rotating trapped BECs have attracted much attention
because of an interesting link \cite%
{Wilkin,Wilkin-CP,Cooper,anyon,Ho,Ho1,spin1,PRA-highL,noBEC2D} of the system
with the quantum Hall effect (QHE) for a two-dimensional (2D) electron gas
in a strong magnetic field \cite{Laughlin,QHE-book,MacDonald}. In these
experiments approaching the \textit{vortex matter} in a rotating BEC, a
large angular momentum can be deposited to the condensate by rotating it at
a frequency close to the quadrupolar resonance. Since BECs are superfluids,
the imprinted angular momentum can only be carried by quantized vortices,
leading to similar physics as in type-II superconductors and quantum Hall
liquids.

In this context, the possible quantum Hall regime of rapidly-rotating BECs (%
\textit{strongly correlated atoms}) is of high interest as it is possibly
the bosonic twin of the QHE for electrons. So far, several interesting ideas 
\cite{Wilkin,Wilkin-CP,Cooper,anyon,Ho,Ho1,spin1,PRA-highL} parallel to the
usual QHE have been explored in this regime. These include, e.g., the
variational Laughlin-like ground states \cite{Wilkin,Cooper}, the concept of
composite particles \cite{Wilkin-CP} and $\frac{1}{2}$-anyons which obey $%
\frac{1}{2}$-statistics \cite{anyon}. In these studies, the profound
connection to the quantum Hall physics, accompanied with exact
diagonalization and variational studies, has given important physical
insights into the various strongly correlated phases of rapidly-rotating
BECs. However, in what sense and to what extent the analogy works are not
clear enough. These questions will determine how far we can proceed into the
possible quantum Hall physics in rapidly-rotating BECs.

In this paper we demonstrate that there is a much deeper connection between
the strongly-correlated bosonic and fermionic systems in their quantum Hall
regimes. We establish a novel map (or correspondence) between the atomic and
electronic QHEs. The map allows one to gain a clearer insight to the
physical meaning of the filling fraction and, more interestingly, to define
the atomic counterpart of the Hall conductance [see Eq.~(\ref{Hall}) below]
that is quantized for atomic Laughlin states. We suggest a vortex analogy by
mapping the atomic Laughlin wavefunctions onto a classical statistical
mechanics problem for 2D vortices \cite{2Dvortex-RMP,Thouless}, resulting in
another useful connection of rapidly-rotating BECs to the 2D vortex physics
which has been extensively studied. We then give an effective Chern-Simons
(CS) theory \cite{Wen-AP} to describe the quantum Hall phase of the present
system. The vortex analogy and the CS theory lead to surprising
predictions---fractionalization of atomic mass and emergence of a
gravitional-like field in atomic quantum Hall liquids.

\textit{Atomic Laughlin states}.---We consider a BEC (with $N$ bosonic atoms
of mass $m$) which is trapped in the $x$-$y$ plane by an isotropic harmonic
potential rotating along $\hat{z}$ at a frequency $\omega $. The BEC can be
effectively treated as a 2D system by assuming the confinement to be strong
enough in the $\hat{z}$-direction \cite{anyon,EPJD}. The single-particle
Hamiltonian describing an atom in the BEC can be written, in a frame of
reference rotating at the frequency $\omega $, as \cite{Cooper,anyon,Ho,EPJD}
\begin{eqnarray}
H^{rot} &=&\frac{\mathbf{p}^{2}}{2m}+\frac{1}{2}m\omega _{0}^{2}\mathbf{r}%
^{2}-\omega \hat{z}\cdot \mathbf{r}\times \mathbf{p}  \nonumber \\
&=&\frac{\left( \mathbf{p}-m\omega _{0}\hat{z}\times \mathbf{r}\right) ^{2}}{%
2m}+(\omega _{0}-\omega )\hat{z}\cdot \mathbf{r}\times \mathbf{p},
\label{ham}
\end{eqnarray}%
where $\mathbf{r}=(x,y)$, $\mathbf{p}=-i\hbar \nabla $, $\mathbf{r}\times 
\mathbf{p}\equiv \hat{z}L_{z}$ is the the angular momentum of the atom, and $%
\omega _{0}$ is the natural frequency of the trap. In the limit $\omega
_{0}=\omega $ (the \textquotedblleft QHE limit\textquotedblright ), $H^{rot}$
is completely analogous \cite{Wilkin,Wilkin-CP,Cooper,anyon,Ho,Ho1} to the
Hamiltonian $\frac{1}{2m_{e}}\left( \mathbf{p}+e\mathbf{A}/c\right) ^{2}$ of
an electron moving in the same geometry subjected to a magnetic field $B\hat{%
z}=\nabla \times \mathbf{A}$ with $\mathbf{A}=\frac{1}{2}B\hat{z}\times 
\mathbf{r}$. The \textquotedblleft lowest Landau level\textquotedblright\
(LLL) states of the atom then reads $\psi _{l}(\eta )=N_{l}\eta ^{l}\exp %
\left[ -\left\vert \eta \right\vert ^{2}/(4\ell ^{2})\right] $ ($%
l=0,1,2,\cdots $) in terms of the complex coordinate $\eta =x+iy\equiv \ell 
\bar{\eta}$. Here $N_{l}=\left[ \pi l!(2\ell ^{2})^{l+1}\right] ^{-1/2}$ is
the normalization constant and $\ell =\sqrt{\hbar /(2m\omega _{0})}$ the
\textquotedblleft magnetic length\textquotedblright , the minimal length
scale in the problem. $\psi _{l}(\eta )$ is also the eigenstate of $L_{z}$
with eigenvalue $l$.

The many-body Hamiltonian of the BEC is $H=H_{0}+H_{L}+H_{g}$ with $%
H_{0}=\sum_{i=1}^{N}\frac{\left( \mathbf{p}_{i}-m\omega _{0}\hat{z}\times 
\mathbf{r}_{i}\right) ^{2}}{2m}$, $H_{L}=(\omega _{0}-\omega )\hat{z}\cdot
\sum_{i=1}^{N}\mathbf{r}_{i}\times \mathbf{p}_{i}$ and $H_{g}=g%
\sum_{i<j=1}^{N}\delta (\mathbf{r}_{i}-\mathbf{r}_{j})$. Here $g=U_{0}\sqrt{%
\frac{m\omega _{z}}{2\pi \hbar }}$ \cite{EPJD} and $U_{0}=4\pi a\hbar ^{2}/m$
measures the strength of the atom--atom interaction in three-dimensional
space; $\omega _{z}$ is the trapping frequency in the $\hat{z}$-direction
and $a$ ($>0$) denotes the $s$-wave scattering length. Now we are concerned
only with the LLL subspace; the energy scales associated with $H_{0}$ and $%
H_{g}$ are assumed to be much large than that of $H_{L}$\ \cite{anyon}. In
this limit, the physics is dominated by the energy scale characterizing $%
H_{g}$.

By analogy to the Laughlin wavefunction for the usual QHE \cite{Laughlin},
one can take the following ansatz of the ground state of the rotating BEC in
the QHE regime 
\begin{equation}
\Psi _{l}=\mathcal{N}_{l}\prod_{i<j}\left( \bar{\eta}_{i}-\bar{\eta}%
_{j}\right) ^{l}e^{-\frac{1}{4}\sum_{i}\left\vert \bar{\eta}_{i}\right\vert
^{2}},  \label{Laughlin}
\end{equation}%
where $\mathcal{N}_{l}$ is an unimportant normalization constant and $l>0$
an even number. The $\frac{1}{2}$-anyons proposed in Ref. \cite{anyon} are
described by $\Psi _{l}$ with $l=2$, which is the simplest atomic Laughlin
state. Experimental conditions for realizing $\Psi _{2}$ were analyzed in
Ref.~\cite{anyon}. A crucial difference between the present atomic system
and the electronic system is that, for the latter the interaction is the
usual (long-range) Coulomb potential, while the atom-atom interaction is the
contact interaction. Actually, it has been shown that the Laughlin state is
the exact nondegenerate ground state for repulsive interactions of vanishing
range \cite{exact-Kiv}. This statement is equally applicable to bosons \cite%
{exact-Xie}. Thus $\Psi _{l}$ is the exact ground state of $H_{0}+H_{g}$ (or 
$H$ if $\omega _{0}=\omega $); when the effect of $H_{L}$ is taken into
account, $\Psi _{2}$ will be the ground state with lowest angular momentum 
\cite{anyon}. It seems to be reasonable to say that \textit{rapidly-rotating
BECs offer a better opportunity for observing the QHE in its bosonic version}%
. Crucially, exact diagonalization studies indicate the chemical potential
discontinuity for $\nu =\frac{1}{2}$, $1$, $\frac{3}{2}$, \ldots , $6$ \cite%
{Cooper}. This implies the incompressibility of the rotating BEC described
by $\Psi _{2}$, a property that is a prerequisite for the emergence of QHE 
\cite{QHE-book,MacDonald}.

\textit{Atomic quantum Hall effect}.---To make the analogy with the usual
QHE more precisely, let us consider again an atom whose motion is determined
by $H^{rot}$ in the QHE limit. Now assume for the time being that in the
rotating frame of reference, the atom is also subjected to a fictitious
gravity $mg^{\ast }\hat{y}$, whose physical meaning will become clear later.
If there is a mass current density $j_{x}=nm\dot{x}$ (with $n$ being the
number density of atoms) driven in the $x$-direction, then at equilibrium,
the fictitious gravity $mg^{\ast }\hat{y}$ balances the Coriolis force $2m%
\mathbf{\dot{r}}\times \omega _{0}\hat{z}=-2m\dot{x}\omega _{0}\hat{y}$.
Similarly, one can define the \textquotedblleft Hall
conductance\textquotedblright\ in this classical consideration as 
\begin{equation}
\sigma _{xy}^{(m)}=\frac{j_{x}}{g^{\ast }}=\frac{nm\dot{x}}{2\dot{x}\omega
_{0}}=\frac{nm}{2\omega _{0}}.  \label{xym}
\end{equation}%
However, for the rotating BEC described by the Laughlin wavefunction (\ref%
{Laughlin}), the Hall conductance $\sigma _{xy}$ so defined is also
predicted, as in the usual QHE, to be quantized as 
\begin{equation}
\sigma _{xy}^{(m)}=\frac{nm}{2\omega _{0}}=\frac{m^{2}}{2\pi \hbar }\nu ,
\label{Hall}
\end{equation}%
with the filling fraction $\nu =1/l$. An interesting observation arising
from Eq.~(\ref{Hall}) and the usual quantized Hall conductance is the fact
that the mass $m$ plays exactly the same role in the atomic QHE as the
charge $e$ in the electronic QHE. This implies a crucial correspondence $%
m\leftrightarrow e$ between atomic and electronic QHEs. The physical meaning
of $\nu $ can be illustrated as what follows. For the condensed atoms spread
over an area $S$, one can interpret the line integral $\int_{\partial
S}\omega _{0}(\hat{z}\times \mathbf{r}_{i})\cdot d\mathbf{l=}2\omega _{0}S$
along the closed boundary $\partial S$ as the total vortices. Note that the
quantum of vorticity (circulation) is $2\pi \hbar /m\equiv \phi _{0}$. By
identifying $\nu $ as the ratio between the atom number and total number of
vortices \cite{Cooper}, it now takes a physically appealing form 
\begin{equation}
\nu =\frac{nS}{2\omega _{0}S/(2\pi \hbar /m)}=\frac{2\pi \hbar }{m^{2}}\frac{%
nm}{2\omega _{0}},  \label{nu}
\end{equation}%
which just gives rise to Eq.~(\ref{Hall}).

Recall that the QHE leads to a high-accuracy resistance standard and
independent determination of the fine-structure constant \cite{QHE-book,vK}.
A striking consequence of the quantization of $\sigma _{xy}^{(m)}$ [Eq.~(\ref%
{Hall})] is that, as far as the quantized $\sigma _{xy}^{(m)}$ can be
actually measured, a high-precision measurement of the fundamental quantity $%
\frac{m^{2}}{2\pi \hbar }$ of atoms is conceivable, which might find
important applications in some other contexts.

\textit{Fractional excitations}.---In the quantum Hall states for a 2D
electron gas, the system has fractionally charged excitations \cite{Laughlin}%
. An important problem in the present context is to identify the excitations
in the atomic quantum Hall states of the rotating BEC. Since the BEC under
study is a macroscopic quantum state of neutral atoms, it cannot support
charged excitations. As done by Laughlin in dealing with the electronic
quantum Hall states, one can similarly map the ground state wavefunction $%
\Psi _{l}(\bar{\eta}_{1},\cdots ,\bar{\eta}_{N})$ onto a classical
statistical mechanics problem by $\left\vert \Psi _{l}(\bar{\eta}_{1},\cdots
,\bar{\eta}_{N})\right\vert ^{2}=e^{-\Phi _{l}/(\kappa _{B}T)}$, where $%
\frac{1}{\kappa _{B}T}=\frac{1}{l}K^{-1}$ (with $\kappa _{B}$ and $T$
standing for the Boltzmann constant and an effective temperature,
respectively) and 
\begin{equation}
\Phi _{l}=-K\sum_{i<j}2l^{2}\ln \left\vert \bar{\eta}_{i}-\bar{\eta}%
_{j}\right\vert +\frac{K}{2}l\sum_{i}\left\vert \bar{\eta}_{i}\right\vert
^{2}.  \label{potential}
\end{equation}%
Remarkably, if one takes $K=\frac{\pi n\hbar ^{2}}{m}$, then the first term
in Eq.~(\ref{potential}) represents precisely the potential energy of $N$
vortices interacting with each other via logarithmic potentials \cite%
{2Dvortex-RMP,Thouless}; each of the vortices has $-l$ vorticity quanta.
Meanwhile, the second term in $\Phi _{l}$ describes the interaction between
the vortices and a uniform neutralizing background of vortex density $\sigma
=1/(2\pi \ell ^{2})$ (Note that $\nabla ^{2}\left[ \frac{1}{2}\left\vert
\eta \right\vert ^{2}\right] =4\pi \sigma $).

Having established the natural \textit{vortex analogy}, instead of the
plasma analogy used first by Laughlin in the usual QHE \cite{Laughlin}, one
is ready to consider the elementary excitations which are created from the
ground state $\Psi _{l}$ by adiabatically inserting (removing) a vorticity
quantum $\phi _{0}$ at position $\bar{\eta}_{0}$ and read 
\begin{eqnarray}
\Psi _{l}^{+}\left( \bar{\eta}_{0}\right) &=&\mathcal{N}_{l}^{+}\prod_{i}%
\left( \bar{\eta}_{i}-\bar{\eta}_{0}\right) \Psi _{l}(\bar{\eta}_{1},\cdots ,%
\bar{\eta}_{N}),  \label{p} \\
\Psi _{l}^{-}\left( \bar{\eta}_{0}\right) &=&\mathcal{N}_{l}^{-}e^{-\frac{1}{%
4}\sum_{i}\left\vert \bar{\eta}_{i}\right\vert ^{2}}\prod_{i}\left( 2\frac{%
\partial }{\partial \bar{\eta}_{i}}-\bar{\eta}_{0}\right)  \nonumber \\
&&\times \prod_{j<k}\left( \bar{\eta}_{j}-\bar{\eta}_{k}\right) ^{l},
\label{n}
\end{eqnarray}%
where $\mathcal{N}_{l}^{\pm }$ are two normalization constants. Generalizing
to the cases of many excitations is straightforward. Following the above
vortex analogy and similarly to Laughlin's argument, a striking result---%
\textit{an exotic fractionalization of atomic mass}---immediately follows: 
\textit{The state }$\Psi _{l}^{+}\left( \bar{\eta}_{0}\right) $\textit{\ }[$%
\Psi _{l}^{-}\left( \bar{\eta}_{0}\right) $]\textit{\ describes an
excitation, or atomic quasihole }(\textit{quasiatom})\textit{, with a
fractional mass }$-m/l$ ($m/l$)\textit{; the fractional mass can even be
negative for atomic quasiholes}. This is very similar to the
fractionalization of the elementary charge in the usual QHE. It is obvious
that fractional excitations obey fractional statistics \cite%
{Laughlin,QHE-book}, e.g., $\frac{1}{2}$-anyons proposed in Ref.~\cite{anyon}
obey $\frac{1}{2}$-statistics. The fractional mass and fractional statistics
can also be obtained more directly and rigorously from the adiabatic theorem 
\cite{Arovas}. Moreover, the composite particles \cite{Wilkin-CP} and
fermionization of bosonic atoms \cite{anyon} naturally arise in the present
theory: Attaching an odd (even) number of vorticity quantum to a bosonic
atom results in a composite fermion (boson); actually, any (bosonic,
fermionic, or fractional) statistics is possible due to the unique property
of 2D space \cite{MacDonald}.

The exotic fractionalization of atomic mass can also be understood by
following Laughlin's thought\textit{\ }experiment \cite%
{Laughlin,QHE-book,MacDonald}. Here the crucial point is that the ground
state is gapped and incompressible. By adiabatically inserting a vorticity
quantum $\phi _{0}$ at the origin in a disk geometry, then a radical mass
density $j_{r}$ is driven out to the boundary and necessarily induces an
azimuthal field $g_{\phi }^{\ast }$ due to the Hall response. Now seen along
a ring (with radius $R$) far away from the origin, the induced mass $\nu m$
is 
\begin{equation}
\nu m=\sigma _{xy}^{(m)}\int \frac{d\phi }{dt}dt=\int dtj_{r}2\pi R=\sigma
_{xy}^{(m)}\int dtg_{\phi }^{\ast }2\pi R,  \label{mmu}
\end{equation}%
where Eqs.~(\ref{xym}) and (\ref{Hall}) have been used. Thus on the one
hand, to be consistent with the fractional mass obtained from the vortex
analogy, Eq.~(\ref{mmu}) implies that $g_{\phi }^{\ast }$ \textit{stems from
the time dependence of vorticity }$\phi $\textit{\ and is given by} 
\begin{equation}
g_{\phi }^{\ast }=\frac{1}{2\pi R}\frac{d\phi }{dt},  \label{gfi}
\end{equation}%
\textit{which is exactly the counterpart of the Faraday induction law}. On
the other hand, one can get the correct fractional mass [Eq.~(\ref{mmu})] if
taking Eq.~(\ref{gfi}) as a starting point. For $\nu =\frac{1}{2}$, the
thought experiment creates at fixed total atom number quasiparticles with
mass $\pm m/2$.

\textit{Cherm-Simons\ effective theory}.---As one can see from the
derivation of Eq.~(\ref{gfi}), $g_{\phi }^{\ast }$ arises from the system
being an incompressible quantum Hall liquid. Now we proceed to show that the
situation is more naturally incorporated by an effective CS theory which has
been exploited successfully in the usual QHE \cite{Wen-AP} and will be
considered here only briefly (For details, see Ref.~\cite{Chen}). Define
covariant three-dimensional spacetime vector $x_{\alpha
}=(x_{1},x_{2},x_{3})=(\mathbf{r},c_{s}t)$ ($\alpha =1$, $2$, $3$; $c_{s}$
could be the speed of sound whose precise value is unimportant here) and the
density vector $J_{\alpha }=(J_{1},J_{2},J_{3})=(\mathbf{J},nc_{s})$. Here $%
t $ is the time coordinate. Then Eqs.~(\ref{xym}), (\ref{Hall}) and (\ref%
{gfi}) imply that the \textit{external} velocity field $V_{\alpha
}=(V_{1},V_{2},V_{3})=(\mathbf{V},V_{3})$ ($\mathbf{V}=\omega _{0}\hat{z}%
\times \mathbf{r}$) will induce the Hall response of the atom number
density: 
\begin{equation}
m\delta J_{\alpha }=c_{s}\sigma _{xy}^{(m)}\varepsilon ^{\alpha \beta \gamma
}\partial _{\beta }\delta V_{\gamma },  \label{jv}
\end{equation}%
where $\varepsilon ^{\alpha \beta \gamma }$ is the totally antisymmetric
unit tensor. We intend to find a Lagrangian to produce Eq.~(\ref{jv}). For
this purpose, one can introduce a $U(1)$ \textit{velocity CS field} $%
v_{\alpha }$ such that $J_{\alpha }=\sum_{i}\dot{x}_{\alpha }\delta (\mathbf{%
r}-\mathbf{r}^{\prime })=\frac{c_{s}}{\phi _{0}}\varepsilon ^{\alpha \beta
\gamma }\partial _{\beta }v_{\gamma }$, which automatically guarantees the
conservation of the density $J_{\alpha }$. By including the source term $%
j_{\alpha }$ of excitations, the complete effective theory is given by the
CS Lagrangian 
\begin{equation}
\mathcal{L}_{k}=\frac{c_{s}}{2\phi _{0}}m\varepsilon ^{\alpha \beta \gamma
}\left( V_{\alpha }-\frac{l}{2}v_{\alpha }\right) \partial _{\beta
}v_{\gamma }+\frac{1}{2}kmv_{\alpha }j_{\alpha }.  \label{lk}
\end{equation}%
Here $k=\pm 1$, $\pm 2$, \ldots ; $k=\pm 1$ correspond to elementary
excitations and $k\neq \pm 1$ to composite excitations.

Now the equations of motion are 
\begin{equation}
mJ_{\alpha }=\nu kmj_{\alpha }+c_{s}\sigma _{xy}^{(m)}\varepsilon ^{\alpha
\beta \gamma }\partial _{\beta }V_{\gamma },  \label{em}
\end{equation}%
where the first (fractional mass $\nu km$) term comes from the increase of
atom density associated with the excitations and the second term gives the
Hall response, e.g., 
\begin{equation}
\left( 
\begin{array}{c}
mJ_{1} \\ 
mJ_{2}%
\end{array}%
\right) =\left( 
\begin{array}{cc}
0 & \sigma _{xy}^{(m)} \\ 
\sigma _{xy}^{(m)} & 0%
\end{array}%
\right) \left( 
\begin{array}{c}
g_{1}^{\ast } \\ 
g_{2}^{\ast }%
\end{array}%
\right) .  \label{hres}
\end{equation}%
The absence of the diagonal elements of the $2\times 2$ matrix in Eq.~(\ref%
{hres}) implies a dissipationless mass current. Here $\mathbf{g}^{\ast
}=c_{s}\nabla V_{3}-\frac{\partial }{\partial t}\mathbf{V}$ is the induced
gravitational-like acceleration and plays the same role as the
current-induced electric field in the usual QHE. Surprisingly, $\mathbf{g}%
^{\ast }$ (or $g_{\phi }^{\ast }$) is nothing but the fictitious
gravitational acceleration introduced previously by hand; here it is not
fictitious at all and, instead, a natural consequence of the
incompressibility of the strongly-correlated atoms. Intuitively, one could
use the following picture: The mass current will experience the Coriolis
force produced by $\mathbf{V}$; the current-induced excitations with
opposite masses then move in opposite directs, inducing the
gravitational-like potential $c_{s}V_{3}$. The role of $\mathbf{V}$ ($%
c_{s}V_{3}$) is thus quite similar to that of the external vector potential $%
\mathbf{A}$ (the current-induced electrostatic potential). The \textit{%
emergence of a gravitational-like field in the atomic quantum Hall liquids}
is of fundamental interest. Though being very amazing and unexpected, it is
an unavoidable consequence of our theory.

As is known in the context of the usual QHE \cite{Wen-AP}, the CS effective
theory is an equivalent description of QHE. The CS theory, as we presented
in this work, is interesting in its own right since it deals with a
non-electromagnetic field in $2+1$ dimensions. Generalization of the simple
Lagrangian (\ref{lk}) is similar to Ref.~\cite{Wen-AP} and might imply rich
topological orders in the atomic quantum Hall liquid. Interestingly, based
on the effective theory the gapless \textit{edge states} \cite{Wen-AP} are
predicted \cite{Chen} to exist in the quantum Hall regime of
rapidly-rotating BECs and represent a chiral Luttinger bosonic liquid; more
importantly, they open up the possibility of detecting the bulk properties
of the system and even $\mathbf{g}^{\ast }$.

To summarize, we have shown a nearly perfect analogy between the atomic QHE
for rotating BECs and the usual QHE, with also, of course, some important
differences which do not alter the overall picture. This profound similarity
stems from the powerful correspondences between the atomic and electronic
QHEs 
\[
m\longleftrightarrow e,\ \ V_{\alpha }\longleftrightarrow A_{\alpha },\ \
v_{\alpha }\longleftrightarrow a_{\alpha }, 
\]%
where $A_{\alpha }$ ($a_{\mu }$) is the external (CS) electromagnetic field 
\cite{Wen-AP}. It allows one to use many existing results developed for the
usual QHE to understand the physics of rotating BECs in the quantum Hall
regime. Our analysis has substantiated the existence of the atomic QHE being
the twin of the electronic QHE as far as the atomic quantum Hall regime is
accessible. In particular, we have predicted the quantization of the atomic
Hall conductance and the exotic fractionalization of atomic mass for atomic
quantum Hall liquids, a new state of matter in atomic gases with unique
strongly-correlated properties. The effective CS theory can be consistently
constructed and makes the same predictions to the microscopic theory. The
realization of atomic quantum Hall liquids is still challenging under
current experimental conditions. Several recent experiments \cite%
{vortex-Science,vortex-critical,vortex-topo} have already made an important
first step to achieve the goal.

\begin{acknowledgments}
This work was supported by the National Natural Science Foundation of China,
the Chinese Academy of Sciences and the National Fundamental Research
Program (under Grant No. 2001CB309303).
\end{acknowledgments}

\end{document}